\documentclass{WileyMSP-template}
\usepackage{pdfpages}

\newcommand{\tsub}{\textsubscript}

\begin{document}
\pagestyle{fancy}
\rhead{\includegraphics[width=2.5cm]{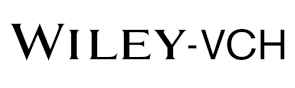}}

\title{Charge-Density Wave Driven Giant Thermionic-Current Switching in 1T-TaS$_{2}$/2H-TaSe$_{2}$/2H-MoS$_{2}$ Heterostructure}
\maketitle

% Author: Please give full first and last names for authors and include * after the name of all corresponding authors
\author{Mehak Mahajan}
\author{Kausik Majumdar*}

% Dedication
%\dedication{}

% Affiliations: Please provide adacemic titles (Prof. or Dr.) for all authors where applicable, and include an institutional email address for all corresponding authors
\begin{affiliations}
Mehak Mahajan, Dr. Kausik Majumdar\\
Department of Electrical Communication Engineering\\
Indian Institute of Science\\
Bangalore 560012, India\\
Email Address: kausikm@iisc.ac.in
\end{affiliations}

% Keywords: Please provide a minimum of three and a maximum of seven keywords, separated by commas

% Abstract should be written in the present tense and impersonal style (i.e., avoid we), and be at most 200 words long
\begin{abstract}
1T-TaS$_2$ exhibits several resistivity phases due to the modulation of charge density wave (CDW). The fact that such phase transition can be driven electrically has attracted a lot of attention in the recent past towards \emph{active-metal} based electronics. However, the bias-driven resistivity switching is not very large ($<$ 5 fold), and an enhancement in the same will highly impact such phase transition devices. One aspect that is often overlooked is that such phase transition is also accompanied by a significant change in the local temperature due to the low thermal conductivity of 1T-TaS$_2$. In this work, we exploit such electrically driven phase transition induced temperature change to promote carriers over a thermionic barrier in a 1T-TaS$_{2}$/2H-TaSe$_{2}$/2H-MoS$_{2}$ T-Junction, achieving a $964$-fold abrupt switching in the current through the MoS$_2$ channel. The device is highly reconfigurable and exhibits an abrupt reduction in current as well when the biasing configuration changes. The results are promising for several electronic applications, including neuromorphic chips, switching, nonlinear devices, and industrial electronics such as current and temperature sensing.
\end{abstract}

% Text: Please use section headings and subheadings as specified below. For communications, all section headings apart from Experimental Section should be removed
% Please make the first reference to a display item bold: \textbf{Figure 1}
% Do not abbreviate Figure, Equation, etc.; display items are always singular, i.e., Figure 1 and 2.
% Equations are always singular, i.e., Equation 1 and 2, and should be inserted using the {equation} environment, not as graphics
% Please do not use footnotes in the text, additional information can be added to the Reference list.

\section{Introduction}
The charge density wave (CDW) driven phase transition in layered materials like  TaS$_2$, TaSe$_2$, NbSe$_2$, and TiSe$_2$ has attracted a lot of attention in the recent past due to their structural, thermal, magnetic, electrical and optical properties \cite{Bana2013,Yan2018,Mahajan2019light,Zhang2020,Li2021}. Out of all these, 1T-TaS$_2$ exhibits polymorphic states with distinctive resistive phases, that can be achieved by thermal \cite{Manzke1989,Thompson1971}, mechanical \cite{Bu2019,Sruthi2021}, optical \cite{Zhu2018,Li2020}, and electrical \cite{Hollander2015,Yoshida2017} excitation. On heating, 1T-TaS$_2$ transits from commensurate (\textit{C}) to triclinic (\textit{T}) phase at $223$ K, \textit{T} to non-commensurate (\textit{NC}) phase at $283$ K, \textit{NC} to incommensurate (\textit{IC}) phase at $353$ K and \textit{IC} to metallic phase at $550$ K. On the other hand, 2H-TaSe$_2$ that undergoes \textit{C} to \textit{NC} phase transition at $90$ K and \textit{NC} to metallic phase at $120$ K shows a slight change in slope of resistance \textit{versus} temperature curve. Unlike 1T-TaS$_2$, 2H-TaSe$_2$ does not exhibit a sharp change in the resistance during the phase transitions.

The fact that in 1T-TaS$_2$, the resistance switching through such phase transitions can be obtained through Joule heating by electrical driving \cite{Hollander2015,Yoshida2017,Mahajan2019}, has led to several electronic and optoelectronic device applications \cite{Yoshida2015,Liu2016,Zhu2018,Mahajan2020}. The resistance switching ratio plays an important role in determining the performance of these devices. However, the degree of the resistance switching is often weak, less than $5$ \cite{Hollander2015,Yoshida2017,Liu2016}, depending on the flake thickness, measurement temperature, and crystal quality. Any technique that can enhance the ratio of the resistance switching will be of great importance in such phase transition-based device applications. In this work, we propose a method using a 1T-TaS$_{2}$/2H-TaSe$_{2}$/2H-MoS$_{2}$ T-junction, where we achieve a gate-controllable current switching ratio up to $964$ driven by CDW phase change in 1T-TaS$_2$ - which is more than $190$-fold enhancement in the switching ratio compared to existing reports. The principle of operation of the technique exploits the abrupt change in the junction temperature due to an abrupt change in the current resulting from Joule heating induced CDW phase transition in TaS$_2$. This, in turn, helps to promote carriers over a thermionic barrier, modulating the net probe current. Interestingly, the proposed device is electrically reconfigurable between a positive jump (enhancement in current) and a negative jump (reduction in current). Such reduction in current can be modeled as an effective negative differential resistance (NDR), as shown earlier \cite{Mahajan2020}.

%\subsubsection{First Sub Subsection}
%\threesubsection{First lowest-level subsection}

\section{Results and Discussion}
\begin{figure}[!h]
  \includegraphics[width=\linewidth]{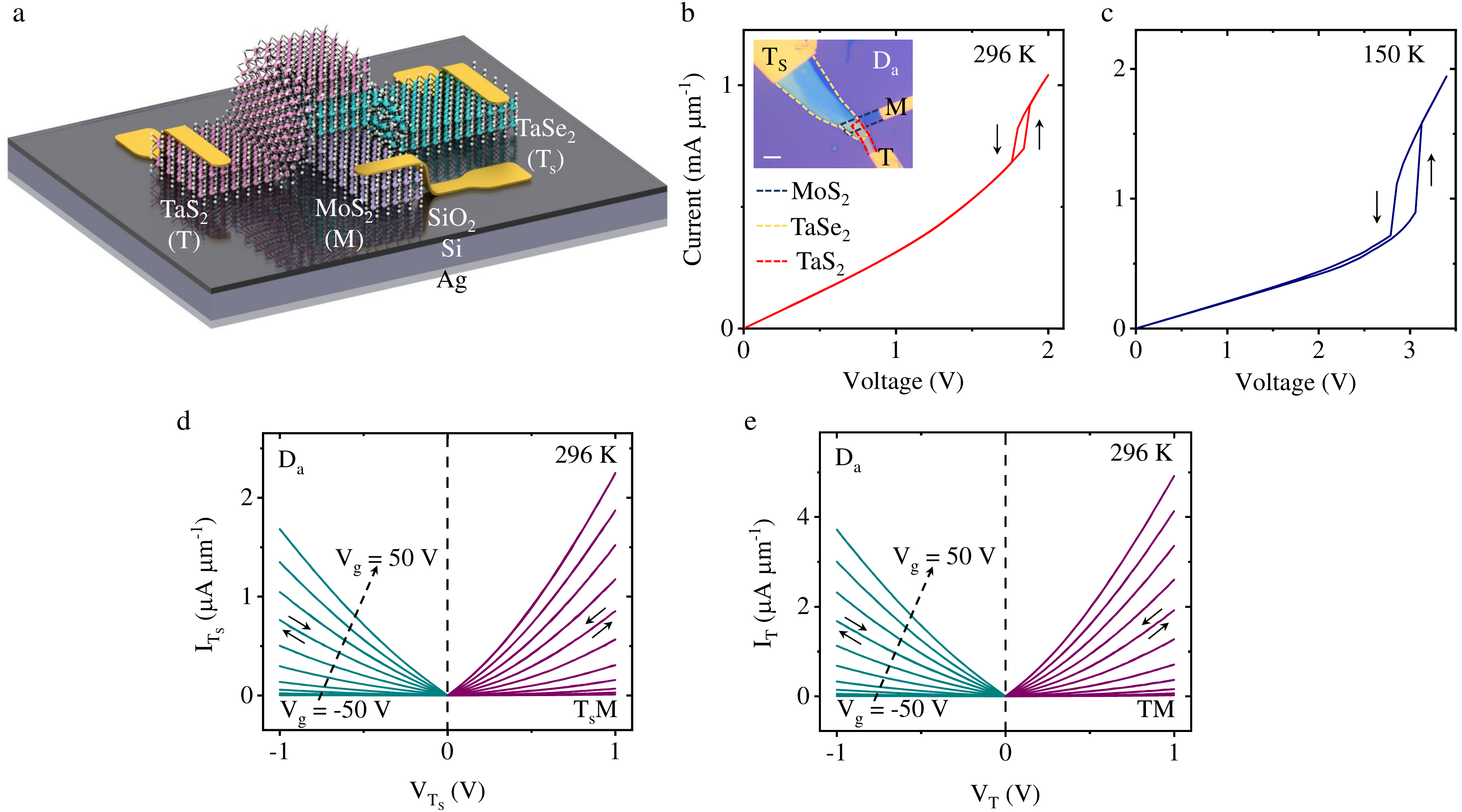}
  \caption{Electrical Characterization of individual materials and heterojunctions of the T-junction. (a) Schematic representation of triple layered T-junction. The Ag contact is used as a global back gate.(b)  Current-voltage characteristics of two-probe TaS$_2$/TaSe$_2$ junction (probed between terminals T\tsub{S} and T of D\tsub{a}) indicating the \textit{NC}-\textit{IC} phase transition of TaS$_2$ at $296$ K. Inset: The optical image delineating each layer of the fabricated triple layered T-junction (D\tsub{a}) (Scale bar: $5\ \mu m$). (c) Current-voltage characteristics of TaS$_2$/TaSe$_2$ junction at $150$ K. Forward and reverse sweeps are indicated by black arrows. (d) I-V characteristics of TaSe$_2$/MoS$_2$ junction, probed between terminals T\tsub{S} and M keeping terminal T open, for $V_g$ varying from $-50$ V to $50$ V in steps of $10$ V at $296$ K. (e) I-V characteristics of TaS$_2$/MoS$_2$ junction, probed between terminals T and M keeping terminal T\tsub{S} open, as a function of $V_g$ varying from $-50$ V to $50$ V in steps of $10$ V at $296$ K.}
  \label{fig1}
\end{figure}

The 3D schematic diagram of the proposed triple layered T-junction is shown in \textbf{Figure \ref{fig1}}a. The inset of Figure \ref{fig1}b shows the optical image of the device D\tsub{a} fabricated by exfoliating a few layer MoS$_2$ flake on Si/SiO$_2$ substrate followed by layer-by-layer dry transfer of TaSe$_2$ and TaS$_2$ flakes respectively (see the experimental section for complete fabrication steps). The current (\textit{I}) - voltage (\textit{V}) characteristics of the TaS$_2$/TaSe$_2$ junction of the device D\tsub{a} probed between terminals T and T\tsub{S} (with terminal M kept open) at $296$ K are shown in Figure \ref{fig1}b. The abrupt hysteretic jump in the current ($\sim 1.24$) results from the reduction in the TaS$_2$ resistance owing to electrically driven \textit{NC}-\textit{IC} CDW phase transition. Note that TaSe$_2$ being highly conductive, carries ample current to drive \textit{NC}-\textit{IC} phase transition of TaS$_2$ in TaS$_2$/TaSe$_2$ junction \cite{Mahajan2019light}. Also, the low thermal conductivity of TaSe$_2$ compared to Au contact further aids in raising the local temperature of the channel \cite{Regueiro1985,Bernd2009}. The current jump can be further enhanced at low temperatures by invoking phase transition through multiple metastable states of TaS$_2$ using the external electric field \cite{Liu2016, Zhu2018}. Figure \ref{fig1}c depicts the \textit{I}-\textit{V} characteristics of the TaS$_2$/TaSe$_2$ junction of the device D\tsub{a} at $150$ K showing a current jump of $\sim 1.77$. Note that when a material with lower conductivity (such as SnSe$_2$) replaces the series resistance material (that is, TaSe$_2$), the overall current is suppressed, which reduces the Joule heating. Hence, the phase transition in TaS$_2$ does not occur anymore (Supplementary Figure S1).

The current-voltage characteristics of TaSe$_2$/MoS$_2$ junction (probed between terminals T\tsub{S}  and M keeping terminal T open) and TaS$_2$/MoS$_2$ junction (probed between terminals T and M keeping terminal T\tsub{S} open) as a function of back-gate voltage ($V_g$) varying from $-50$ V to $50$ V in steps of $10$ V at $296$ K are outlined in Figure \ref{fig1}d and \ref{fig1}e respectively. The forward and reverse sweep directions of the voltage during measurement (indicated by the black arrows in Figure \ref{fig1}d-e) show negligible hysteresis, suggesting excellent interface quality in the device. The corresponding gate-dependent current-voltage characteristics of both the junctions at $150$ K are depicted in supplementary Figure S2a and S2b respectively.

\subsection{Reconfigurable abrupt current switching in T-junction}
\begin{figure}[!h]
  \includegraphics[width=\linewidth]{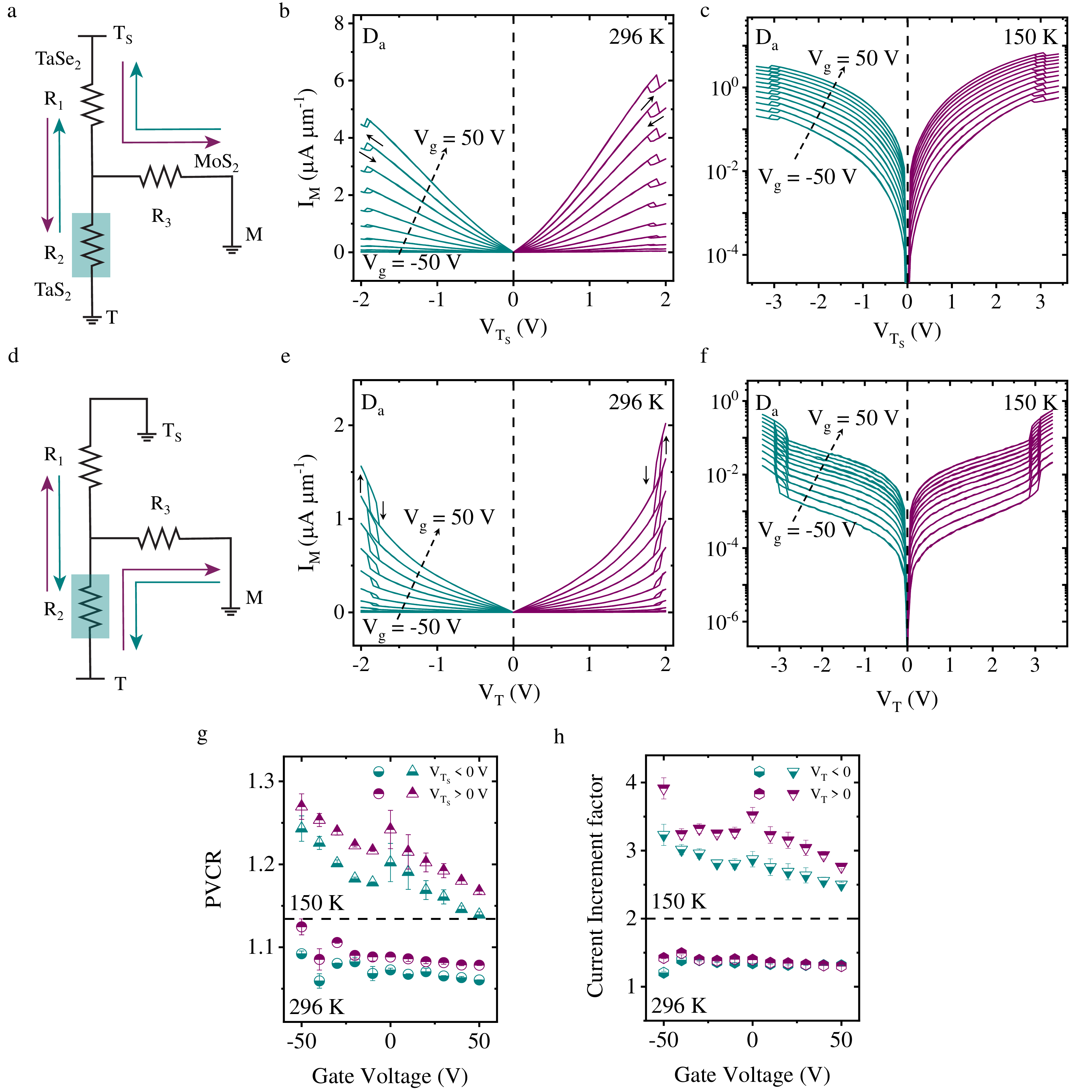}
  \caption{Gate-tunable, reconfigurable abrupt current increment and negative differential resistance. (a),(d) The three branches of the device D\tsub{a} depicted by three effective resistances R$_1$, R$_2$ and R$_3$, corresponding to TaSe$_2$, TaS$_2$ and MoS$_2$, respectively, for TaSe$_2$ (T) biasing [in (a)] and TaS$_2$ (T\tsub{S}) biasing [in (d)]. R$_3$ includes the MoS$_2$ branch resistance and the TaSe$_2$/MoS$_2$ Schottky junction resistance. The cyan box (in the TaS$_2$ branch) indicates the branch where CDW phase transition occurs. The arrows indicate the direction of the current flow in different biasing configurations. (b),(c) Current through the MoS$_2$ branch (I\tsub{M}) \textit{versus} V\tsub{T\tsub{S}} as a function of $V_g$ varying from $-50$ V to $50$ V in steps of $10$ V at $296$ K and $150$ K depicting current decrement for both V\tsub{T\tsub{S}} $ > 0$ and V\tsub{T\tsub{S}} $ < 0$. Forward and reverse sweeps are indicated by black arrows. (e),(f) I\tsub{M} \textit{versus} V\tsub{T} as a function of $V_g$ varying from $-50$ V to $50$ V in steps of $10$ V at $296$ K and $150$ K depicting abrupt current increment for both V\tsub{T} $ > 0$ and V\tsub{T} $ < 0$. Forward and reverse sweeps are indicated by black arrows. (g) Peak-to-valley current ratio (PVCR) as the function of $V_g$ at $296$ K and $150$ K in bottom and top panel respectively. (h) Current increment factor as a function of $V_g$ at $296$ K and $150$ K in bottom and top panel respectively.}
  \label{fig2}
\end{figure}

We now show the characteristics of 1T-TaS$_{2}$/2H-TaSe$_{2}$/2H-MoS$_{2}$ T-junction that is reconfigurable between abrupt current increment and decrement by utilizing the modulation of the MoS$_2$ resistance through an external gate voltage. In this device, we pass a high current through a voltage bias through the TaS$_2$/TaSe$_2$ path and use the branched out MoS$_2$ channel current as a probe. The equivalent circuit diagram of the four terminal device D\tsub{a} is shown in \textbf{Figure \ref{fig2}}a wherein R$_1$, R$_2$ and R$_3$ correspond to the effective resistance of TaSe$_2$, TaS$_2$ and MoS$_2$ flakes respectively. Here, R$_3$ includes both the TaSe$_2$/MoS$_2$ Schottky junction resistance and the MoS$_2$ channel resistance, and both the components are tunable by the gate voltage. A global back gate is connected to the fourth terminal of the device. Figure \ref{fig2}b shows the variation of MoS$_2$ channel current (I\tsub{M}) with V\tsub{T\tsub{S}} of the device D\tsub{a} when we apply the bias at T\tsub{S} terminal while keeping T and M terminals grounded. The characteristics show current decrement behaviour for both V\tsub{T\tsub{S}} $< 0$ V and V\tsub{T\tsub{S}} $> 0$ V for $V_g$ varying from $-50$ V to $50$ V in steps of $10$ V at $296$ K. When the external bias exceeds the threshold voltage for the \textit{NC}-\textit{IC} phase transition of TaS$_2$, the TaS$_2$ resistance abruptly reduces resulting in the abrupt increment in the TaS\tsub2 current, and hence a simultaneous reduction in the MoS\tsub2 current. The peak-to-valley current ratio (PVCR) obtained for V\tsub{T\tsub{S}} $< 0$ V and V\tsub{T\tsub{S}} $> 0$ V at $296$ K are shown in bottom panel of Figure \ref{fig2}g. A simplistic way to further improve the PVCR values is by increasing the switching ratio of the TaS$_2$ resistance, which directly regulates the abrupt change in the MoS$_2$ current. A base temperature lower than the \textit{C}-\textit{T} phase transition temperature can be utilized to achieve such enhancement. We operate the device (D\tsub{a}) at a temperature below the \textit{C}-\textit{T} transition temperature to invoke field-driven metastable states of TaS$_2$ that increase the abrupt current jump by $1.42\times$ in comparison to room temperature. Figure \ref{fig2}c depicts I\tsub{M} \textit{versus} V\tsub{T\tsub{S}} characteristics delineating current decrement at $150$ K. The corresponding PVCR values at $150$ K are shown in the top panel of Figure \ref{fig2}g reaching a maximum value of $1.27$ at $V_g$ $= -50$ V for V\tsub{T\tsub{S}} $> 0$ V which is higher in comparison to the room temperature values.

Suppose we apply the bias at the TaS$_2$ terminal while keeping terminals T\tsub{S} and M grounded as depicted in the equivalent circuit in Figure \ref{fig2}d. In that case, the MoS$_2$ current exhibits an abrupt increment instead of a decrement at the phase transition. When we apply a high electric field across the TaS$_2$/TaSe$_2$ junction, the TaS$_2$ phase transition occurs, which reduces the resistance of the TaS\tsub2 branch resulting in the abrupt increment in the TaS\tsub2 current as well as the MoS\tsub2 current. Here, the total current flow through R$_2$ instead of R$_1$ (in the case of TaSe$_2$ biasing) as shown by the arrows in Figure \ref{fig2}d. Figure \ref{fig2}e and \ref{fig2}f shows the gate voltage ($V_g$) dependent current increment characteristics for both V\tsub{T} $ < 0$ V and V\tsub{T} $ > 0$ V at $296$ K and $150$ K, respectively. The corresponding current increment factor as the function of $V_g$ for both $296$ K and $150$ K is shown in Figure \ref{fig2}h. Current increment factor reaches a maximum value of $\sim 4$ at $150$ K for V\tsub{T} $> 0$ V which is $\sim 2.75 \times$ higher than the room temperature value.

The low-temperature current increment factor of $4$ cannot solely result from phase transition induced resistance switching of TaS$_2$ as it is limited to a factor of $\sim 1.77$ as shown in Figure \ref{fig1}c. In addition to 1T-TaS$_2$, 2H-TaSe$_2$ also exhibits CDW phase transitions at $90$ K and $120$ K. However, it does not cause any abrupt discontinuity in the TaSe$_2$ resistance apart from a slight slope change in resistance \textit{versus} temperature curve \cite{Lee1970,Naito1982,Mahajan2019light}. The overall change in resistance of the TaSe$_2$ branch from the estimated temperature values (discussed later) is small and rules out the contribution of TaSe$_2$ resistance change to the current jump. Also, the heat equation solution of the channel and the in-situ Raman measurement exclude the possibility of structural phase transition of MoS$_2$ from the semiconducting (2H) to metallic (1T) phase. The abrupt current increment due to the TaS$_2$ phase transition increases the local temperature along the device channel. The one-dimensional heat equation solution (using FEM) for the TaS$_2$/TaSe$_2$ channel estimates that the local temperature increment due to Joule heating is sufficient to invoke TaS$_2$ phase transitions (see Supplementary Figure S3), depending upon the base temperature as well as the external bias. However, it is not enough to induce structural 2H to 1T phase transition in MoS$_2$ \cite{Hwang2018}. This argument is further supported by bias-dependent Raman measurement (see Supplementary Figure S4), wherein j-peaks have not been observed, which could otherwise confirm the presence of the 1T-phase of MoS$_2$.

\begin{figure}[!h]
  \includegraphics[width=\linewidth]{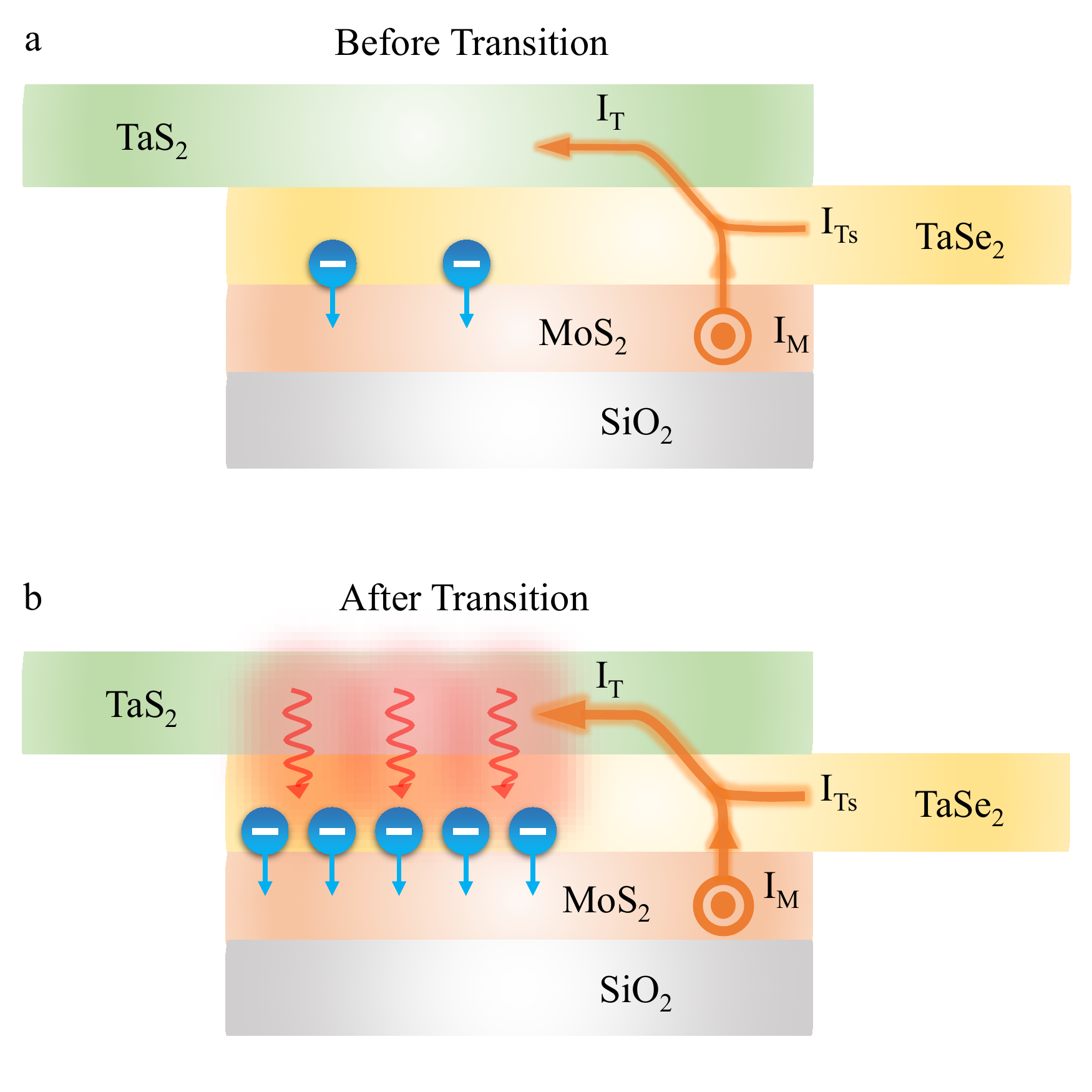}
  \caption{Schematic illustration of the mechanism. Schematic representation of the situation (a) before and (b) after the CDW phase transition in TaS$_2$. The orange arrow and the out-of-plane direction symbol indicate the direction of the current flow though the junction and in the MoS$_2$ channel, respectively. The heat flow from TaS$_2$ to the TaSe$_2$/MoS$_2$ junction in (b) is indicated by the red curved arrows. The abrupt enhancement in temperature promotes a larger number of electrons through the TaSe$_2$/MoS$_2$ Schottky barrier, in turn causing an abrupt change in the MoS$_2$ channel current.}
  \label{schematic}
\end{figure}

Such a large current increment can be explained by the enhanced thermionic carrier injection across the TaSe$_2$/MoS$_2$ barrier with a sudden increase in the junction temperature resulting from the Joule heating induced phase transitions of TaS$_2$, as detailed next. The mechanism is schematically illustrated in Figure \ref{schematic}a-b. The abrupt jump in temperature due to the CDW phase transition in TaS$_2$ causes a corresponding increment in temperature at the TaSe$_2$/MoS$_2$ junction due to efficient heat conduction through the TaS$_2$/TaSe$_2$ interface. That, in turn, enhances the kinetic energy of the carriers, helping them overcome the TaSe$_2$/MoS$_2$ Schottky barrier.

\subsection{Giant current increment driven by thermionic switching}
\begin{figure}[!h]
	\includegraphics[width=\linewidth]{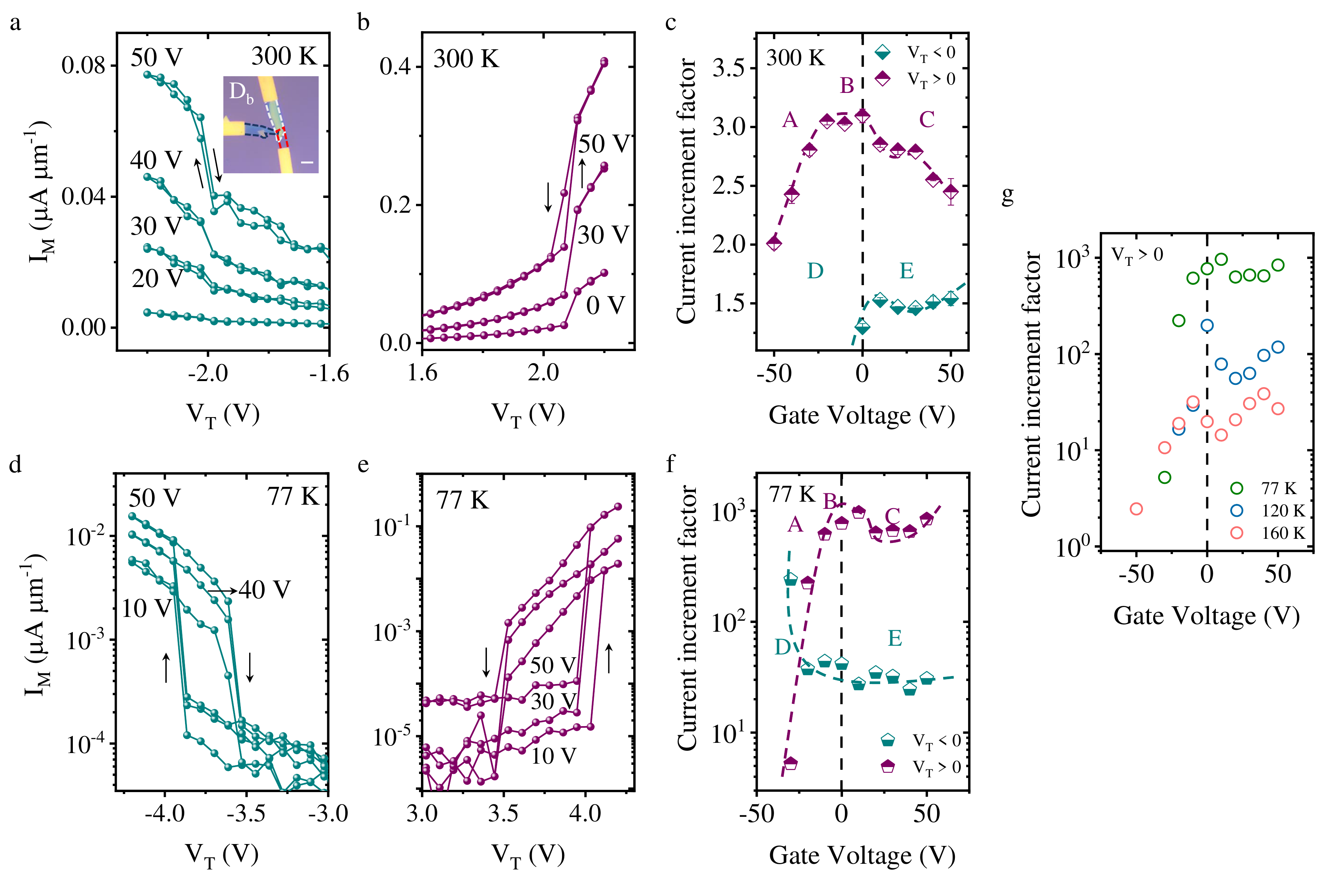}
	\caption{Giant current increment in triple layered junction. (a), (b) I$_M$ \textit{versus} V\tsub{T} for selected $V_g$ at $300$ K for V\tsub{T} $<0$ [in (a)] and V\tsub{T} $>0$ [in (b)] depicting abrupt current increment. Forward and reverse sweeps are indicated by black arrows. The inset of (a) shows the optical image of the fabricated triple layered T-junction (D\tsub{b}) (Scale bar: $5\ \mu m$). (d),(e) I$_M$ \textit{versus} V\tsub{T} for selected $V_g$ at $77$ K for V\tsub{T} $<0$ [in (d)] and V\tsub{T} $>0$ [in (e)] depicting abrupt current increment. (c),(f) Current increment factor as the function of $V_g$ at $300$ K [in (c)] and $77$ K [in (f)] extracted from (a),(b) and (d),(e) respectively. (g) Current increment factor as a function of $V_g$ at $77$, $120$ and $160$ K.}
	\label{fig3}
\end{figure}

To explore the thermionic effect due to local temperature switching, we fabricate another triple layered T-junction (D\tsub{b}) comprising a narrow MoS$_2$ channel with a reduced junction overlap area (see the optical image in the inset of \textbf{Figure \ref{fig3}}a). Here we choose a small junction overlap area between MoS$_2$ and the TaS$_2$/TaSe$_2$ to further increase the temperature by forcing the hot electrons to pass through a smaller region. Apart from weaker heat dissipation efficiency, the smaller junction area helps in two other ways. First, when the junction area is small, the fractional contribution of the junction resistance to the total device resistance increases. That helps to achieve a larger switching ratio since the hot carriers, due to the phase transition-induced temperature change, primarily modulate the junction resistance. Second, a smaller overlap area helps to remove any trapped residue and air from the interface during annealing, which in turn helps to achieve closer proximity between MoS$_2$ and TaSe$_2$/TaS$_2$ heater, helping the MoS$_2$ layer to achieve a higher temperature effect at the junction during the switching. I\tsub{M} \textit{versus} V\tsub{T} of device D\tsub{b} for selective $V_g$ values varying from $20$ V to $50$ V (step size: $10$ V) at $300$ K for V\tsub{T} $< 0$ V is depicted in Figure \ref{fig3}a. Similarly, Figure \ref{fig3}b shows the I\tsub{M} \textit{versus} V\tsub{T} for V\tsub{T} $> 0$ at $V_g$ equal to $0$ V, $30$ V and $50$ V. The corresponding current increment factors during the phase transition are plotted in Figure \ref{fig3}c showing a maximum value of $\sim 3$ at $V_g$ $= -10$ V for V\tsub{T} $> 0$ V at $300$ K which is $2 \times$ higher in comparison to the current increment factor of device D\tsub{a}.

At low temperature, we could further enhance the current jump as depicted in I\tsub{M} \textit{versus} V\tsub{T} characteristics of device D\tsub{b} at $77$ K (see Figure \ref{fig3}d and \ref{fig3}e for V\tsub{T} $< 0$ V and V\tsub{T} $> 0$ V, respectively) for different $V_g$ values. The corresponding current increment factors are plotted in Figure \ref{fig3}f for both V\tsub{T} $< 0$ V and V\tsub{T} $> 0$ V. We could enhance the MoS$_2$ current by a factor of $\sim 964$ at $V_g$ $= 10$ V for V\tsub{T} $> 0$ V as represented in Figure \ref{fig3}f. As discussed earlier, such a huge jump can not solely arise from electrically driven TaS$_2$ phase transition. And, we must invoke the corresponding sudden rise in the junction temperature due to the abrupt enhancement in current through TaS$_2$/TaSe$_2$ during the CDW phase transition of TaS$_2$. Figure \ref{fig3}g depicts the variation of the current increment factor as a function of $V_g$ at a base temperature of $77$ K, $120$ K, and $160$ K for V\tsub{T} $> 0$ V. Such a strong temperature dependence in the current increment factor suggests possible usage of the technique in sensing temperature.

\begin{figure}[!h]
	\includegraphics[width=\linewidth]{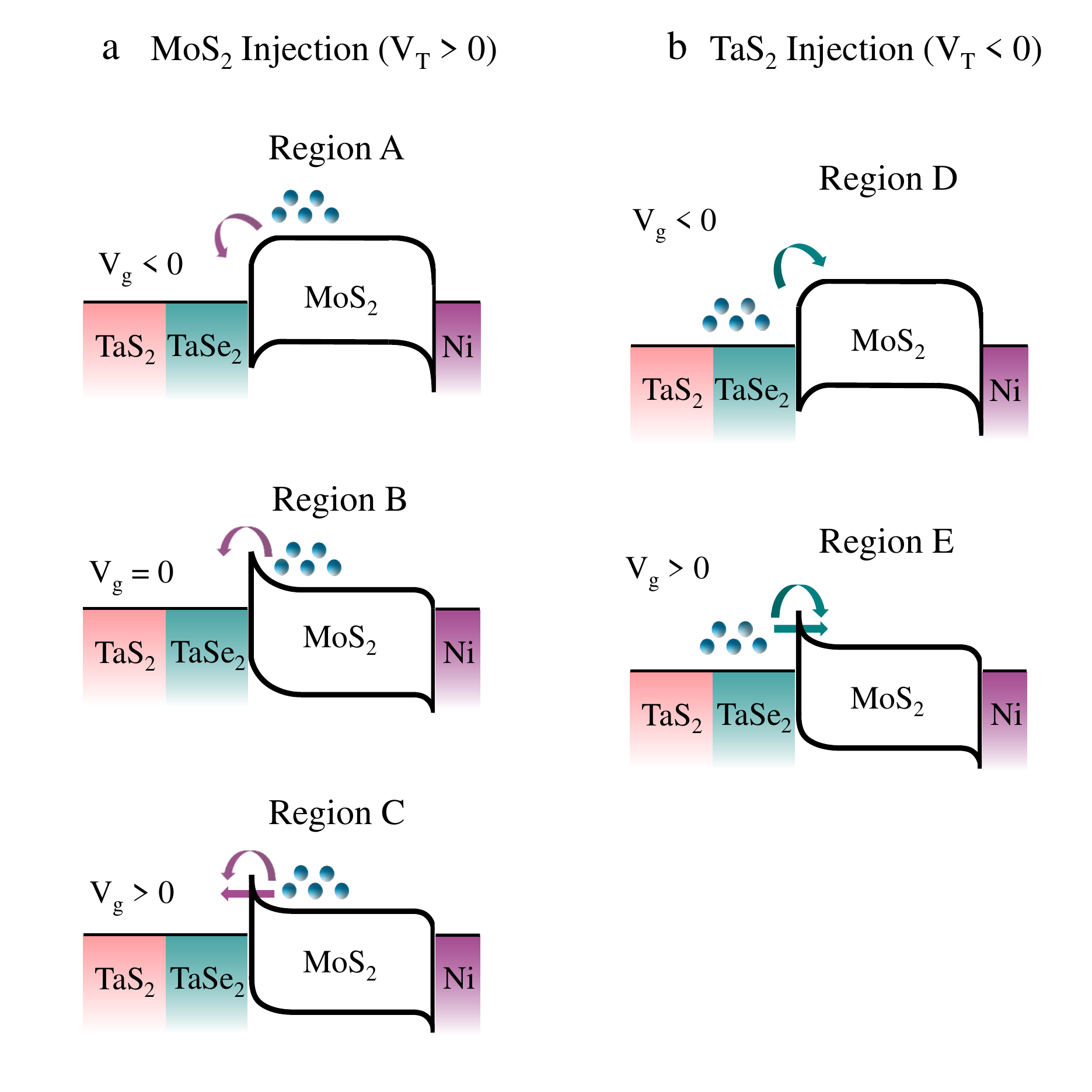}
	\caption{Thermionic switching behavior. (a) Band alignment schematics for V\tsub{T} $ > 0$, that is, when MoS$_2$ injects electron into TaSe$_2$ through the drain barrier at $V_g < 0$ (in top panel), $V_g = 0$ (in middle panel), and $V_g > 0$ (in bottom panel) corresponding to region A, B, and C in figure \ref{fig3}. (b) Band diagram for TaSe$_2$ injection into MoS$_2$ through the source barrier (V\tsub{T} $< 0$) for $V_g < 0$ (in top panel), and $V_g > 0$ (in bottom panel) illustrating region D and E from Figure \ref{fig3}c and f.}
	\label{fig4}
\end{figure}

In Figure \ref{fig3}f for $V_T>0$, the current jump ratio at the phase transition exhibits a strong non-monotonic behavior with $V_g$. Also, the ratio shows an opposite trend for $V_T<0$, particularly when $V_g <0$. The origin of such behavior is explained in \textbf{Figure \ref{fig4}}. In the case of V\tsub{T} $> 0$ V, that is, when MoS$_2$ injects electrons into TaSe$_2$, there are three different regions of device operation (viz. A, B, and C as shown in Figure \ref{fig3}c and \ref{fig3}f) as schematically explained in Figure \ref{fig4}a. Note that, due to the high electrical conductivity of TaSe$_2$ in comparison to TaS$_2$ (in particular, at a low temperature), the floating voltage at the triple junction is small; hence the drop across the MoS$_2$ channel is also small (estimated to be $< 26$ mV). Due to such a small effective drain voltage, the current through the MoS$_2$ channel strongly depends on the drain barrier as well in the different regimes of operation. In the current situation, where TaSe$_2$ acts as the drain contact for $V_T>0$ configuration, the drain barrier plays an even more important role. We recently found that due to van der Waals nature of the contact interface, TaSe$_2$ exhibits strong Fermi level depinning \cite{Murali2021} with layered semiconductors. Accordingly, due to the relatively large work function of TaSe$_2$, the conduction band offset between MoS$_2$ and TaSe$_2$ is large. On the other hand, due to Fermi level pinning, Ni/MoS$_2$ junction has a relatively small Schottky barrier height \cite{Somvanshi2017}. That leads to a larger drain barrier compared to the source barrier at the small effective drain bias as schematically shown in Figure \ref{fig4}a, and hence the TaSe$_2$/MoS$_2$ interface controls the drain current.

In region A, that is, for $V_g$ $\ll 0$ V, electrons do not see any barrier at the drain side and can quickly transfer into TaSe$_2$ (see top panel of Figure \ref{fig4}a). Thus, the current does not change much with the abrupt increase in temperature during the CDW phase transition. In region B, the electrons from the MoS$_2$ region require to overcome the drain barrier to be transferred to TaSe$_2$ (see middle panel of Figure \ref{fig4}a), and thus the current is a strong function of the local temperature at the drain barrier. The current induced Joule heating can increase the local temperature to $\sim$ $230$ K for a base temperature of $77$ K (see Supplementary Figure S3 for simulated results). The current jump ratio simulated for various barrier heights ($\phi_b$  varying from $30$ to $80$ meV) by solving modified Richardson's equation \cite{Somvanshi2017} at temperatures of $120$ K, $150$ K, $180$ K, $210$ K and $240$ K with respect to $77$ K is shown in Supplementary Figure S5. The simulation results clearly show that the current ratio before and after the phase transition can go as high as $10^4$ depending upon $\phi_b$. That justifies the high current increment factor at $V_g$ $= 10$ V. Finally, in region C for $V_g$ $\gg 10$ V, the band bending increases, and the electrons can tunnel through the Schottky drain barrier, as shown in the bottom panel of Figure \ref{fig4}a. Thus, the local temperature does not affect the current, reducing the current ratio before and after the phase transition. Beyond region C, at an even higher positive $V_g$, we observe an increment in the current increment factor, the origin of which is yet not very clear and could result from a $V_g$ dependent change in the relative resistance between the source and the drain barriers.

On the other hand, for V\tsub{T} $< 0$ V, TaSe$_2$ acts as a source contact, and the electrons are injected from TaSe$_2$ to the MoS$_2$ channel. At high negative $V_g$, i.e., region D, the carrier injection is determined by thermionic transport over the Schottky barrier height at TaSe$_2$/MoS$_2$ interface, as shown in the top panel of Figure \ref{fig4}b, and hence contributes to higher current jump. However, for $V_g$ $> 0$ V (region E), the current increment ratio decreases monotonically as the carrier injection is dominated by the tunneling phenomenon schematically shown in the bottom panel of Figure \ref{fig4}b.

In similarity to device D\tsub{a}, the I\tsub{M} \textit{versus} V\tsub{T\tsub{S}} characteristics of device D\tsub{b} exhibits current decrement for V\tsub{T\tsub{S}} biasing at both $300$ K and $77$ K (see Supplementary Figure S6). The corresponding I\tsub{M} \textit{versus} V\tsub{T} (depicting current increment) and I\tsub{M} \textit{versus} V\tsub{T\tsub{S}} (depicting current decrement) characteristics for device D\tsub{b} at $120$ K and $160$ K are outlined in Supplementary Figure S7 and S8 respectively. The current increment characteristics have been repeatedly observed in several devices, some of which are outlined in Supplementary Figure S9 and S10.

\section{Conclusion}
The technique proposed in this work demonstrates a unique way of significantly amplifying the resistance switching ratio typically obtained from a TaS$_2$ CDW phase transition. That is achieved by exploiting enhanced carrier injection through a Schottky barrier height by the abrupt increment in the local temperature during the phase transition. Accordingly, the technique can be applied to enhance the device performance in several applications where TaS$_2$ phase transition is used, for example, in detecting infrared photons and neuromorphic applications. In addition, the gate tunable sharp jump in current can be useful for sensing applications, such as temperature and current. On the other hand, the enhancement in the MoS$_2$ channel current during the CDW phase transition of TaS$_2$ provides an excellent probe to monitor the local temperature of TaS$_2$.

% Experimental section
\section{Experimental Section}
\threesubsection{Triple Layered T-Junction Device Fabrication and characterization.}
1T-TaS$_{2}$/2H-TaSe$_{2}$/2H-MoS$_{2}$ T-junction is fabricated in the following manner. First, the thin flakes of MoS$_2$ are mechanically exfoliated on a heavily doped Si substrate coated with $285$ nm thick SiO$_2$ using polydimethylsiloxane (PDMS), followed by dry transfer of TaSe$_2$ and TaS$_2$ flakes respectively. A rotational stage controls the alignment during each layer transfer to form the T-junction. The complete exfoliation and dry transfer processes are done at room temperature. The substrate is then spin-coated with a high contrast positive resist $-$ polymethyl methacrylate (PMMA) $950$ C$3$ and softly baked for $2$ minutes at $180$ $^\circ$C. Patterns are formed through electron beam lithography with an electron beam dose of $200$ $\mu$C cm$^{-2}$, an electron beam current of $300$ pA, and an acceleration voltage of $20$ KV. The pattern development is carried out in $1:3$ MIBK/IPA developer solution followed by IPA wash and blow drying in N$_2$. Metal contacts are formed by blanket deposition of $10$ nm Ni / $50$ nm Au using a DC magnetron sputter coating system in the presence of Ar plasma at $ 6.5\times 10^{-3} $ Torr. Excess metal lift-off is carried out by immersing the substrate in acetone for $ 15-30 $ minutes, followed by IPA wash for $30$ seconds and blow drying in N$_2$. Buffered HF solution is used to etch the back oxide from the substrate, and highly conducting silver paste is used for the back gate contact.

The electrical measurements are carried out in a probe station with a base vacuum level of about $1.6 \times 10^{-3}$ Torr at room temperature and $6.45 \times 10^{-6}$ Torr at the low temperature with the supply of liquid N$_2$.

\medskip
\textbf{Supporting Information} \par %Please delete the Suppporting Information statement if it is not applicable. Please supply Supporting Information in another file. Supporting information should not be provided in .tex format
Supporting Information is available from the Wiley Online Library or from the author.

% Acknowledgements
\medskip
\textbf{Acknowledgements} \par %delete if not applicable))
This work was supported in part by a Core Research Grant from the Science and Engineering Research Board (SERB) under Department of Science and Technology (DST), a grant from Indian Space Research Organization (ISRO), a grant from MHRD under STARS, and a grant from MHRD, MeitY and DST Nano Mission through NNetRA.

%References
\medskip
% Use the following code if you wish to generate your bibliography with BibTeX;
% replace the string "MSP-template" below with the name(s) of
% the BibTeX data base(s) you want to use.
% The resulting bibliography-output (the content of the .bbl file)
% must be pasted back into this file before submission.
% Please also include your BibTeX data base file(s) in your submission
% so that we can re-run BibTeX if necessary.
%
\bibliographystyle{MSP}
\bibliography{TJ}

% Table of contents entry should be 50 - 60 words long
% Image should be 55 mm broad and 50 mm high or 110 mm broad and 20 mm high

\begin{figure}[!ht]
\textbf{Table of Contents}\\
\medskip
  \includegraphics[width=7in]{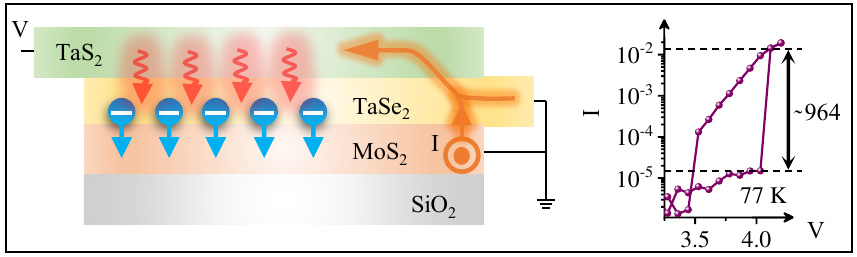}
  \medskip
  \caption*{The electrically driven phase transition of 1\textit{T}-TaS$_2$ and the low thermal conductivity of 2\textit{H}-TaSe$_2$  are exploited in a local heater structure to achieve a $964$-fold abrupt current enhancement through the MoS$_2$ channel in a 1T-TaS$_{2}$/2H-TaSe$_{2}$/2H-MoS$_{2}$ T-junction. The device is highly reconfigurable and exhibits a sharp reduction in current when the biasing configuration changes. In addition, the gate tunability of the current jump makes it useful for various sensing applications, including current and temperature sensors.}
\end{figure}
%\newpage
%\AtEndDocument{\includepdf[pages={1-10}]{supporting_info}}
\end{document}